# Thermal processing and enthalpy storage of an amorphous solid: a molecular dynamics study

P. M. Derlet

*Condensded Matter Theory, Paul Scherrer Institut, PSI-Villigen 5232, Switzerland*

R. Maaß

*Department of Materials Science and Engineering, University of Illinois at Urbana Champaign, 1304 West Green Street, Urbana, Illinois 61801, USA*

**Abstract**

Using very long molecular dynamics simulation runs, temperature protocols spanning up to five orders of magnitude in time-scales are performed to investigate thermally activated structural relaxation in a model amorphous solid. The simulations demonstrate significant local structural excitations as a function of increasing temperature and show that enthalpy rather than energy is primarily responsible for relaxation. The observed enthalpy changes are of the order seen in experiment, and can be correlated with the level of internal hydrostatic stress homogenization and icosahedral content within the solid.

# 1 Introduction

The amorphous solid, such as bulk metallic and network glasses, constitutes a class of materials which attract both scientific and engineering interest [1, 2]. Despite their out-of-equilibrium nature, these materials have impressive elastic (in terms of extension) and plastic (in terms of strength) properties. However, their ductility upon yield is limited because of the emergence of local plastic instabilities, leading to immediate failure under tension and the formation of a few dominant shear bands under compression [3, 4].

Two widely used material preparation protocols employed to influence the mechanical properties of the amorphous solid are that of ageing and rejuvenation [5]. Ageing involves careful thermal annealing procedures which induce structural relaxation and a reduction in energy, resulting in improved elastic properties and a higher yield, but at the expense of ductility. Rejuvenation generally involves structural excitation and an increase in energy, tending to delocalize plasticity and increase ductility, but at the expense of a well defined yield point and thermal stability.

Recently, a number of quite different rejuvenation protocols have been investigated, in particular temperature quenching/ramping between ambient temperatures (well below the glass transition temperature) and liquid Nitrogen temperatures [6], and stress relaxation/creep at ambient temperatures under strains and stresses which are almost two orders of magnitude below the yield point [7]. The implication of these works is that the underlying structural excitations, which collectively lead to plasticity, must be thermally driven or driven by thermal-stresses that develop during cooling.

The amorphous atomic structure contains no discrete translational symmetry, and (so far) no canonical defect framework has been identified which would mediate microscopic plastic processes. Despite this situation, a number of generic atomic scale mechanisms have been proposed using both theory and atomistic simulation. The most widely known mechanisms are those of free-volume migration [8], local shear transformation [9] and shear transformation zone theory [10, 11]. All rely on the concept of a local change in atomic structure, which via the Eshelby inclusion picture [12], results in a far-field plastic strain or stress. Via elasticity, such structural transformations interact leading to complex collective behaviour.

Via simulations, the emergent behaviour of such unit-scale processes is able to rationalize the transition from a low temperature heterogeneous plasticity to a high temperature homogeneous plasticity. This has been seen in thermally driven coarse grained models [13, 14, 15, 16, 17, 18] and stress driven atomistic simulation [10, 19, 20, 21] (see also the review article Ref. [22]). Since the atomistic simulations must be performed at high strain rates, the type of plasticity normally being studied is predominantly athermal. Here the material is driven by stress to a local structural instability. Whilst supporting the athermal effective temperature theories associated with the shear transformation zone construct [11], such plasticity is quite different from the coarse grained models and that of early theory [8, 9] — all of which rely on thermal fluctuations to drive local structural transformations.

The present atomistic simulations will demonstrate that significant structural change under zero load, and well below the glass transition temperature regime, does occur via thermally activated processes when physical simulation times reach the order of several hundred nano-seconds to a micro-second.

In thermally activated plasticity, microscopic plastic processes are driven through thermal fluctuation, where the system passes from one local energy minimum to another via a saddle-point configuration. Here, the external and internal stress biases those processes that relieve the applied load or the internal stress. It is this latter bias that leads to collective and possibly local macro-scopic plasticity. Since the thermal energy scale is small, such unit plastic processes are spatially local and rare when compared to the Debye time-scale. Emergent collective behaviour of many structural transformations will therefore occur at much longer time-scales — many orders of magnitude larger than that probed by atomistic simulation — but comparable to the time scale of a typical deformation experiment. Experimental evidence of thermally activated plasticity is numerous for temperatures approaching the glass transition regime [1] and also at much lower temperatures in shear band nucleation, propagation and arrest [23, 24, 25, 26].

Atomistic simulation techniques able to probe local structural transitions not driven to an instability by an external stress fall into the broad category of potential energy landscape (PEL) exploration algorithms. For model glass systems, the activated-relaxation-technique *nouveau* [27, 28, 29] (ART*n*) has provided important insight into local changes in atomic structure as the system passes through a saddle-point to a new nearby local energy minimum [30, 31, 32, 33, 34, 35, 36, 37]. All these works have shown that model glass structures admit a distribution of barrier energies. When the glass configuration is sufficiently relaxed, this distribution tends to small values for increasingly small barrier energies. The associated structural excitations are local but complex in structure, containing a central group of atoms which undergo a maximum displacement surrounded by a field of much smaller displacements which help to accommodate the central structural change. In refs. [34, 37] these central structures, referred to as local structural excitations, were characterized by a chain-like geometry in which a sequence of neighbouring atoms shifted to adjacent positions. In the work of Fan *et al* [35, 36], such displacements could be correlated quadratically with the corresponding barrier energy suggesting a direct link to the Eshelby inclusion construct [12]. In the work of Swayamjyoti *et al* [37], the barrier energy (and how it changes upon application of an external stress) could be correlated with the stress signature of the corresponding saddle-point configuration under zero-load.

The present work investigates thermally activated local structural excitations under zero load conditions for a model Lennard Jones (LJ) binary structural glass. This is done by studying the degree of increasing atomic scale displacements as a function of increasing temperature using the molecular dynamics method. Four orders of temperature ramping rates are considered, with the slowest rate demonstrating significant activity at temperatures well below the glass transition regime. Sec.II details the simulation methodology and sample preparation methods, Sec.III presents the results of temperature ramping protocols in which the degree of structural relaxation is quantified in terms of quench protocol and temperature ramp rate. Sec.IV investigates the atomic-scale nature of the observed structural excitations and the local environment in which they occur. Using up to a micro-second of physical simulation time, the final section, Sec.V, discusses the results in terms of enthalpy storage and heterogeneous internal pressure distribution, demonstrating that through thermal activation alone, enthalpy changes comparable to experiment can be achieved via atomistic simulation.

## 2 Simulation methodology and glass preparation procedure

Three-dimensional model 50:50 binary glass samples are considered using the Wahnstrøm parametrized Lennard-Jones pair potential [38]. For this potential, the parameter ε sets the energy scale and the parameter $\sigma_{11}$ sets the length scale. The time scale is set by the quantity $\sigma_{11}\sqrt{(m_1/\varepsilon)}$ where $m_1$ and $m_2$ are the masses of the two atom types. For the present simulations the masses of two atom types are arbitrarily chosen to be $m_1$=2 and $m_2$=1 atomic masses. The absolute temperature, *T*, is expressed as an energy, $k_B T$, using $k_B$=8.617×10$^{-5}$ε. When obtained energies are compared to experiment, a characteristic value of ε=1 eV is used. The LAMMPS software [39] is used to perform the molecular dynamics simulations.

A simulation cell consisting of 32000 atoms is considered. Such a size is sufficient to study local thermally activated structural excitations, each event of which can involve up to several atoms.

A very common protocol to produce an amorphous solid at 0K is to begin with a high temperature equilibrium liquid configuration using either a NVT (fixed atomic number, volume and temperature) or NPT (fixed atomic number, pressure and temperature) ensemble. This is then followed by a linear quench in temperature. For the NPT ensemble, a simultaneous pressure quench is also included, such that, as the temperature approaches zero, so does the pressure. If the NVT ensemble is used with a not too slow quench rate, the final sample will generally contain a non-negligible hydrostatic stress component. In the present work, an NVT quench protocol is chosen. Temperature control is performed by rescaling atomic velocities at a chosen time interval. When needed, for a given configuration of atomic positions at a given temperature, displacements due to thermal vibrations are removed by performing a conjugate gradient (CG) relaxation to a zero force configuration. These configurations are referred to as CG-quenched (CG-q) samples.

## 3 Temperature ramping

Fig. 1a displays the potential energy versus time for a linear quench protocol with a temperature quench rate equal to $Q^{LQ}=-6\times10^{-4}$ ε/τ. This corresponds to a drop in temperature equal to $100k_B$ every 10000 molecular dynamics iterations. The potential energy values are derived from an average at each temperature. Two, approximately linear, regimes may be identified with cross-over behaviour in a narrow temperature interval identified as the glass transition temperature regime. Linear extrapolation of the low and high temperature regimes gives a fictive glass temperature equal to $k_B T_G=0.55\varepsilon$. Below this temperature scale, the structure can be considered to be in an amorphous state, whereas above, it may be considered to be in a liquid-like or under-cooled liquid state.

Fig. 1b plots the percentage of atoms which are icosahedrally coordinated as a function of temperature for both instantaneous and CG-q configurations at a given temperature. In both cases, the icosahedral content increases with decreasing temperature. For the case of the CG-q configurations an approximate plateau is reached just below $k_B T_G$, whereas for the instantaneous configurations the increase is more gradual. This difference is explained by instantaneous configurations containing thermal noise in the atomic displacements, which perturb for a short period of time the local structure such that the algorithm fails to identify the icosahedral geometry. The CG-q data of Fig. 1b demonstrates the icosahedral fraction saturates below the temperature regime of $T_G$. Inspection of icosahedral clusters (as defined by a connected set of nearest neighbour icosahedrally coordinated atoms) of this sample, reveals this saturation is not the result of a system spanning cluster, but is rather due to a dense population of clusters consisting of several tens of atoms. Comparison of those atoms icosahedrally coordinated atoms at 0K with those at higher temperatures reveal an overlap greater than 90% for $T<T_G$. Above $T_G$ this overlap drops rapidly to ~10%. Thus the developing icosahedral network effectively locks in (at the timescale of the current quench simulation) for temperatures well below the glass transition.

The final zero temperature sample, referred to as the quenched sample, is now ramped in temperature up to and beyond $T_G$ using four ramp rates spanning four orders of magnitude — $Q_1$, $Q_0$, $Q_{-1}$ and $Q_{-2}$ where $Q_n = -10^n Q^{LQ}$. Fig. 2a displays, as solid lines, the internal energy as a function of increasing temperature for the quenched sample. The data of Fig. 1a is also included as a finely dotted line.

The data reveals little difference in the internal energies between different ramp rates up to a temperature of approximately $0.6 T_G$ — all following the trend associated with the linear quench protocol. However, at higher temperatures, the curves begin to diverge, with the slower ramp rates showing a reduced gradient. This indicates relaxation occurs. At $T_G$, the gradient with respect to temperature of the internal energy increases for all ramp rates, with all curves eventually rejoining the linear quench curve of the under-cooled liquid regime. Inspection of the glass transition regime demonstrates the slower temperature ramp rates transit to the "liquid" state at lower temperatures and over a narrower temperature interval. Thus the slower ramp rates have sufficient time to forget their lower temperature amorphous structure over a shorter temperature interval, just as slower quench rates have sufficient time to forget their higher temperature "liquid" state.

The derivative with respect to temperature of the curves in Fig. 2a gives the heat capacity at constant volume which is measured experimentally for glasses using differential scanning calorimetry (DSC) [5, 42]. Fig. 2b plots the resulting heat capacity for the slowest $Q_{-2}$, and demonstrates a form that follows quite closely what is seen experimentally for metallic glasses. The vertical scale is given in units of J/mol-K and demonstrates (upon integration up to $T_G$) the total heat flow in terms of internal energy is of the order of ~300 J/mol — a result which is comparable to experiment despite the calculation not taking into account total enthalpy flow. These aspects will be discussed in more detail in Sec.V.

To directly observe the degree of relaxation occurring during the temperature ramps, CG-q configuration quantities as a function of temperature are now inspected. Fig. 3 displays the potential energy and icosahedral percentage for the CG-q samples as a function of increasing temperature, for the four considered ramp rates. It is seen that when the ramp rate is lower than the initial quench rate, the energy of the CG-q samples decreases as the temperature is increased. This trend continues to just below $T_G$, after which the CG-q energy rapidly increases, converging to the values of the faster ramp rates. With the observed energy relaxation close to but less than $T_G$, comes an increase in icosahedral content — see Fig. 3b. Thus for the slower ramp rates, the amorphous structure begins to relax well below $T_G$, both in terms of a decreasing cohesive energy and increased icosahedral percentage. The trend of increasing icosahedral content with decreasing energy is a general feature of metallic amorphous systems [40] with past atomistic simulations observing a linear relationship between the two quantities [41].

Together these results indicate significant structural relaxation occurs at temperatures close to, but below, the glass transition regime. This suggests an efficient protocol to produce a relaxed amorphous configuration should involve halting the linear quench protocol at the temperature where this maximum relaxation occurs. At this fixed temperature, the system is then relaxed as long as possible, followed by the normal linear quench protocol towards 0K. This protocol is now performed by taking an atomic configuration at a temperature of $T=0.95T_G$, and allowing it to relax at that temperature for 27 million molecular dynamics iterations. During this relaxation significant energy relaxation and icosahedral fraction increase are observed (not shown). After this, the linear quench protocol is performed at the quench rate $Q^{LQ}$, bringing the sample to zero Kelvin. This sample is referred to as the quench-relaxed sample. Fig. 1a displays the resulting internal energy per atom as a function of temperature for the final linear quench protocol. It is seen that the energy per atom has non-negligibly reduced indicating a more relaxed configuration, whilst the gradient with respect to temperature is constant and equal to that of the quenched sample at low temperatures. Fig. 1b displays the resulting CG-q icosahedral percentage during the linear quench part showing the icosahedral content for this more relaxed sample has increased by approximately three percent.

A similar temperature ramping protocol is performed for the quench-relaxed sample using the four ramp rates $Q_1$, $Q_0$, $Q_{-1}$ and $Q_{-2}$ and the resulting data is shown in Figs. 2 and 3 as dashed lines. The internal energy per atom as a function of increasing temperature (Fig. 2a) demonstrates little difference in internal energies between different ramp rates up to a temperature of approximately $0.95T_G$, indicating much less relaxation has occurred when compared to the quenched sample. At higher temperatures, the faster ramp rates converge to the under-cooled liquid curve at much higher temperatures than for the quenched sample, again indicating the quench-relaxed sample is significantly more relaxed. This is further seen in the CG-q configurations (Fig. 3) which shows similar trends to that seen for the quenched sample, in terms of increasing icosahedral content with decreasing energy, but at a significantly reduced level. These trends are also reflected in the calculated heat capacity (Fig. 2b) which demonstrates a reduced signature at and below the glass transition regime.

## 4 Atomic-scale activity during temperature ramping

To investigate the nature of atomic scale activity as the temperature is ramped, the atomic displacements between the CG-q configurations (used in Fig. 3) of neighbouring temperatures are calculated. This has the advantage of removing thermal noise and explicitly considers displacements associated with a change to a different local potential energy minimum. Such displacement data is used to quantify the level and nature of structural change as a function of temperature and ramp rate.

Direct visual inspection of the atomic displacement fields reveals:

1. atomic displacements may be characterized by a central region containing a number of atoms with maximum displacement magnitude, surrounded by a field of smaller displacements whose magnitudes decrease irregularly with increasing distance from the central region.

2. the frequency of these local displacement structures increases with temperature.

3. in some cases, the displacement structure disappears at a later time, and often at a higher temperature indicating reversible activity. Reappearance is also observed.

4. at the lowest temperatures ($T<0.5T_G$) the atomic displacements of the central atoms are typically a fraction of a bond length, whereas at temperatures approaching $T_G$, the central atomic displacement magnitude is typically a bond length.

Fig. 4a gives an example of a displacement structure occurring at $0.17T_G$, showing the central atoms undergoing the largest displacements, surrounded by a smaller field of accommodating displacements. Only atoms with displacements greater than $0.1\sigma_{11}$ are shown with the largest displacements being $\simeq 0.4\sigma_{11}$. Fig. 4b plots a displacement map of the entire simulation cell at the temperature $0.48T_G$. Here, only atoms with displacement magnitudes greater then $0.5\sigma_{11}$ are shown. Inspection of the figure reveals several local displacement fields in the form of sequential chain like motion in which one atom replaces the position of another atom. Because of the relatively large minimum displacement threshold, the surrounding (and smaller) displacement fields are not seen in this figure. Close inspection of each chain-like structure reveals atomic displacement magnitudes of the order of $\sigma_{11}$, which is the typical length scale of the nearest neighbour distance (see Fig. 3a of Ref. [44]). These displacement structures involve chains of linear extent, strongly curved chains and also closed loop chains. Such string-like atomic displacements have been observed in simulations of the under-cooled liquid [45, 46, 47, 48, 49] and amorphous solid [50, 51, 52, 53] regimes. Moreover, the displacement fields seen in the present dynamical simulations are similar to those obtained using the ART$n$ potential energy landscape exploration method [34, 37]. Following these later past works, such activity will be referred to as local structural excitations (LSEs).

The larger displacements seen in Fig. 4b at the temperature $0.48T_G$, compared to those in Fig. 4 at the temperature $0.17T_G$ reflect the general trend that with increasing temperature, individual atomic displacements increase. This trend has been seen in past work [51, 52] and is compatible with the quadratic relationship between maximum atomic displacement and barrier energy found by Fan *et al* [35]. The displacement structures shown in Figs. 4a and b are from the quench-relaxed sample for the $Q_{-1}$ ramp rate. Similar results are obtained for the other ramp rates and also the quenched sample. This behaviour reflects a transition from a regime of sub-bond-length displacements that might be identified as the so-called rattling-in-the-cage to a regime of super-bond-length displacements [54] — a transition which here occurs below the glass transition regime.

Through an analysis of the characteristic environment and type of the those atoms involved in the localized structural excitations, it is found that atoms of type 2 (the smaller atoms) are mainly involved. For both the quench and quench-relaxed samples, the percentage of atoms involved which are initially icosahedral coordinated is about 2-4% — a percentage significantly lower than the total average seen in Fig. 1b. This is related to the observation that those atoms involved are generally under-coordinated with an average coordination (nearest neighbour number) typically in the range of 9-11.5. For the quenched sample, the average atomic volume of those atoms involved is slightly below that of the global average of type 2 atoms, whereas for the quench-relaxed sample, the average volume is about 20% lower at $T<0.1T_G$ increasing to the volumes seen in the quenched sample at $T\simeq 0.8T_G$. For the quenched sample, the average number of atoms involved in an LSE ranges between 1 and 3, whereas the quench-relaxed sample it is 1-2 atoms. The the more relaxed sample tends to admit more localized LESs — a result also seen in the work of Fan *et al* [35].

## 5 Discussion and concluding remarks

The main findings of the current work may be summarized as:

1. when quenching from the high-temperature equilibrium liquid state, the icosahedral content increases as a function of decreasing temperature until the glass transition temperature regime. Below this temperature regime, and at the time-scale of the quench simulation, the icosahedral content plateaus to a maximum value.

2. efficient relaxation of computer generated amorphous structure occurs at a temperature just below the fictive glass transition temperature.

3. linear temperature ramping protocols, involving four orders of magnitude difference in ramp rate, reveal significant thermally activated relaxation (increased icosahedral content) well below the glass transition temperature.

4. this structural relaxation is mediated by thermally activated local structural excitations (LSEs) consisting of a chain of central atoms undergoing maximum displacement surrounded by a field of small displacements. At lower temperatures, the central displacements are less than that of the mean atomic distance, however, as the temperature approaches the glass transition regime, the central maximum displacements are typically the magnitude of bond length. This suggests a transition from correlated rattling-in-a-cage to that of atomic hops. The temperature regime at which this change occurs will depend on the temperature ramp rate and more generally on the time-interval of observation.

5. the atoms involved in such LSE activity originate in regions characterized mainly by low coordination, high density and low icosahedral content. Atoms of type 2, the smaller atoms, are predominantly involved.

What drives such structural relaxation? Since the simulations were performed at fixed volume, a state variable (in addition to internal energy) will be the stress of the simulation cell. Fig. 5 displays the components of the global stress for the quenched and quench-relaxed samples as a function of temperature for the slowest ramp rate $Q_{-2}$. It is seen that the diagonal components of the quenched sample are non-zero and positive at 0K, and therefore the simulation cell is under a state of compression. Upon temperature ramping, these components reduce almost to zero as $T_G$ is approached, whereas above the glass transition temperature regime they increase as the liquid regime is entered. On the other hand, the quench-relaxed sample has initially negative diagonal components and is therefore under tension. Upon ramping, these values continue to decrease. Together these results suggest that further structural relaxation will result in an increasingly negative (tensile) hydrostatic pressure.

Inspection of Fig. 5 reveals that the off-diagonal stress components remain small relative to the diagonal stress components at low temperatures. This is most likely due the high temperature liquid (from which the glass is prepared) being unable to support a shear. For the quench-relaxed sample, these components fluctuate around zero, whereas for the quenched sample, two components are initially non-negligibly finite at zero temperature. This is probably due to a fluctuation in shear stress of the liquid being frozen-in as the system transits to the amorphous solid regime. Upon temperature ramping these latter shears reduce and eventually fluctuate around zero at temperatures well below the glass transition. Together, these observations indicate it is the reduction in enthalpy (here at fixed volume) which drives structural relaxation, where both the internal energy and hydrostatic pressure are intimately related to atomic structure. In other words, the degree of structural relaxation (here defined by the fraction of icosahedral atoms) correlates most strongly with a reduction in enthalpy at fixed volume.

The above conclusion may be further verified by considering the behaviour of a long-time relaxation simulation at a constant temperature of $0.95T_G$. Fig. 6a plots the instantaneous potential energy over a period of one billion molecular dynamics simulations, which (for a metal) would correspond to approximately one micro-second of physical time. At various time intervals of the relaxation simulation a linear quench at the rate $Q^{LQ}$ is performed to obtain a zero temperature amorphous structure. Fig. 6b plots the resulting potential energy per atom (in units of kJ/mol) as a function of icosahedral content. A similar linear relationship between energy and icosahedral content was seen in Ref. [41]. Also shown is the enthalpy as a function of icosahedral content. Fig. 6b demonstrates the dominance of enthalpy rather than energy as the key parameter, in measuring the energy release/storage of (in this case) structural relaxation.

The high enthalpy release seen in Fig. 6a due to relaxation is comparable to that typically seen in experiments performed at much lower temperatures and longer time-scales than the micro-second considered here [5]. For rapidly quenched (melt-spun) metallic glasses values of up to 5 kJ/mol have been reported [55] with the representative value being about 1 kJ/mol — a lower bound value which has been associated with the creation of a single shear band [56]. The above numbers indicate that the change-in-enthalpy scale in Fig. 6b represents the, so far, maximum attained enthalpy change (energy storage) in a metallic glass. It however should be emphasized that in addition to the high temperature relaxation performed here, the simulated samples should be considered to be in a considerably unrelaxed state when compared to experiment.

The origin of the pressure evolution may be seen via inspection of the local atomic pressure. Fig. 7 displays local pressure histograms taken from the left most ($10^5$ MD steps) and right most ($10^9$ MD steps) configurations used in Fig. 6b. The local pressures are given in units of pressure times volume, and when added together and divided by the total volume of the simulation cell will give the global pressure in units of Pascal. Representing the local pressure in this way avoids the need for defining a local atomic volume for each atom. The data shows that pressure reduction during relaxation at $T=0.95T_G$ involves both a shift and narrowing of the local pressure distribution. Thus, in terms of local pressure, the more relaxed the sample is the more homogeneous the sample becomes at the atomic scale level.

The strong connection between pressure and icosahedral content may be qualitatively rationalized via the premise that the system attempts to maximize local packing via the minimization of bond energies [57]. In three dimensions, this is done via tetrahedral groupings of four neighbouring atoms, which in turn are optimally packed locally to produce the icosahedral structure. Since this cannot be done with each bond energy being locally minimized (even for a mono-atomic system), bond frustration occurs, and a local non-zero pressure develops. Overall structural relaxation may then be seen as a global optimization of this local structure which reflects a decrease in pressure and an increase in icosahedral content as seen in Figs. 6 and 7.

In conclusion, the present simulation work demonstrates significant atomic scale activity well below the glass transition temperature regime for a model Lennard-Jones binary glass system. The microscopic relaxation processes are seen to be mediated by thermally activated localized structural excitations, which at the lower temperature regime takes the form of sub-bond-length atomic displacements. At higher temperatures, which begin to approach the glass transition regime, bond-length atomic displacements increasingly occur. These involve sequential string-like motion in which one atom replaces that of a neighbouring atom. The observed relaxation, as characterized by the icosahedral content, is found to be driven mainly by pressure and therefore enthalpy in which the local atomic pressure distribution narrows with increasing relaxation. Through very long-time simulations involving an order of a micro-second of physical time, experimentally realistic relaxation enthalpies in the kJ/mole range are observed. Whilst the present work has been done under zero-load conditions, the results indicate thermally activated plasticity will occur for arbitrarily small loads — a conclusion which is compatible with recent thermo-plastic experimental bulk metallic glass preparation protocols and which will be investigated in future simulation work.

## 6  Acknowledgements

The authors thank K. Albe, T. Brink, J. Loeffler, H. Roesner, D. Rodney and G. Wilde for insightful discussions.

**Figures**

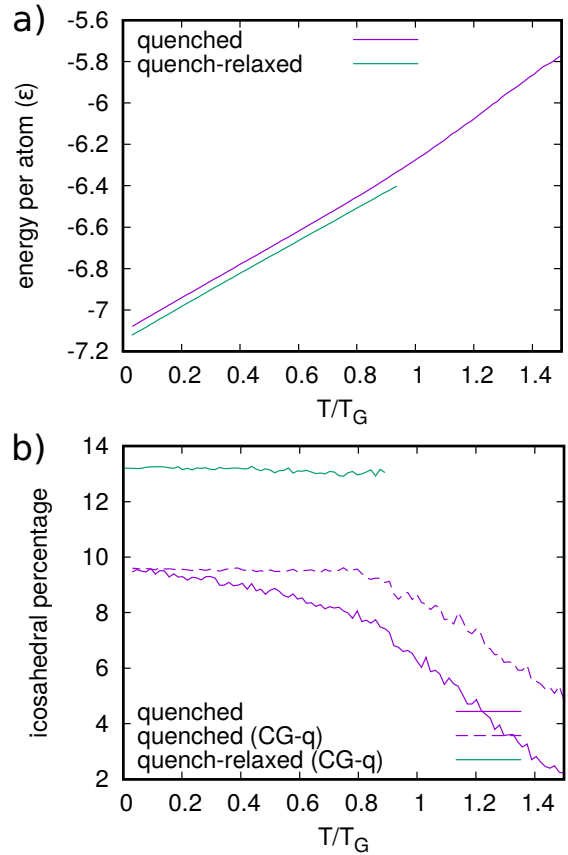

Figure 1: a) Potential energy per atom and b) icosahedral percentage as a function of decreasing temperature of the quenched and the quench-relaxed samples. The quenched curve involves a direct linear quench from a high temperature equilibrium liquid configuration, whereas the quench-relaxed curve is interrupted by a fixed temperature relaxation protocol just below the glass transition regime. For the latter, only the subsequent linear quench data is shown. In b) the icosahedral percentage is also shown for the CG-q configurations which removes the effect of thermal vibrations.

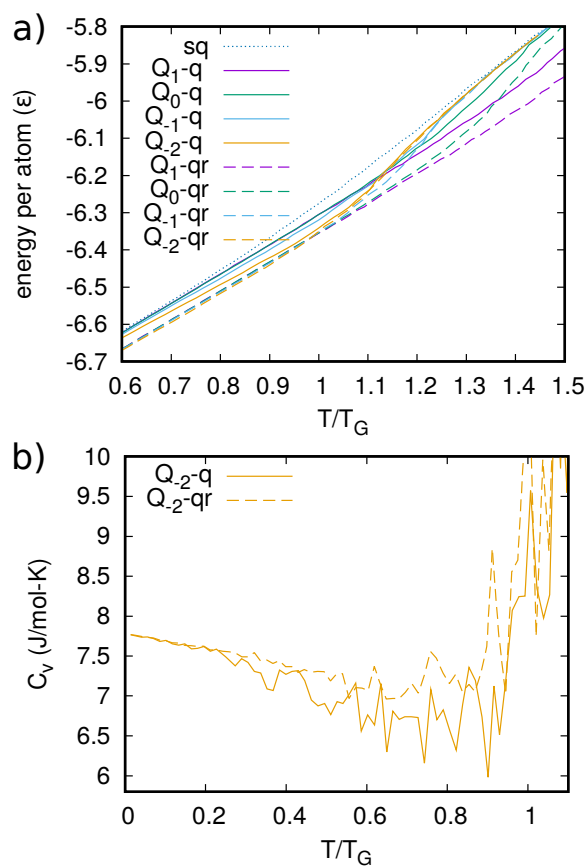

Figure 2: a) Linear ramping protocol from a 0K amorphous structure showing the potential energy per atom as a function of temperature in the vicinity of the glass transition. Data is shown for the four considered ramp rates for both the quenched and quench-relaxed samples. b) The derivative of the potential energy of the slowest ramp rate for the quenched sample, giving the heat capacity at constant volume.

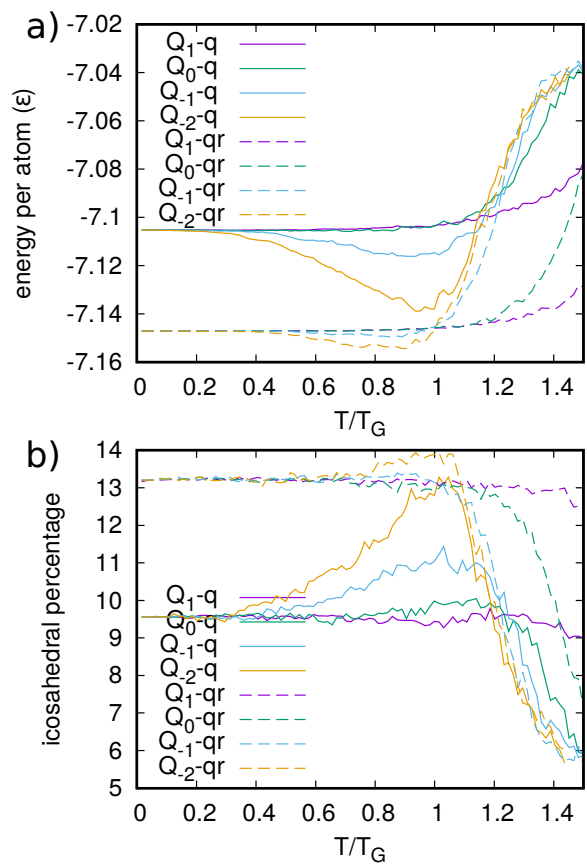

Figure 3: Linear temperature ramping protocol from a 0K amorphous structure showing the CG-q a) potential energy per atom and b) icosahedral fraction as a function of temperature. Data is shown for the four considered ramp rates.

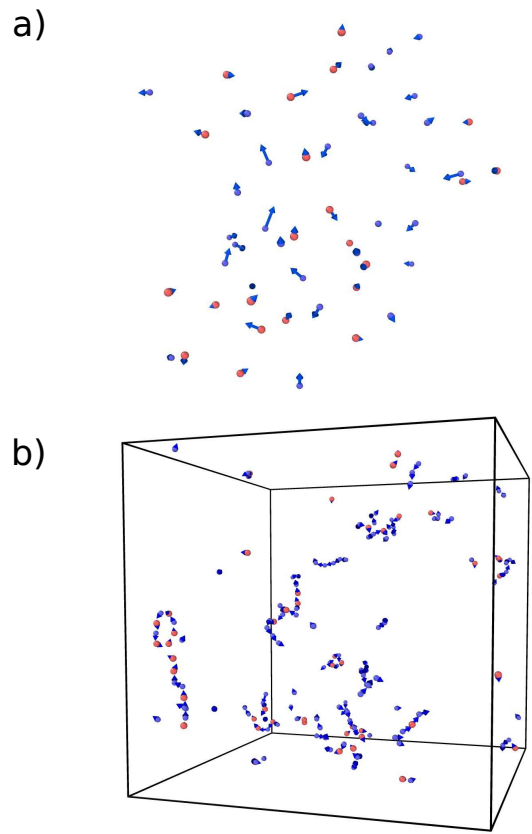

Figure 4: Atomic displacement map at a) the temperature $T=0.17T_G$, showing one local group of atoms undergoing a structural transformation and b) at the temperature $0.48T_G$, showing several local groups of atoms undergoing a structural transformation. Here each vector indicates the atomic displacement and each ball (red/blue corresponding to atoms of type 1/2) indicates the initial position of the associated atom. Only atoms whose displacement is greater than $0.1\sigma_{11}$ are shown. Visualization is via OVITO [43]

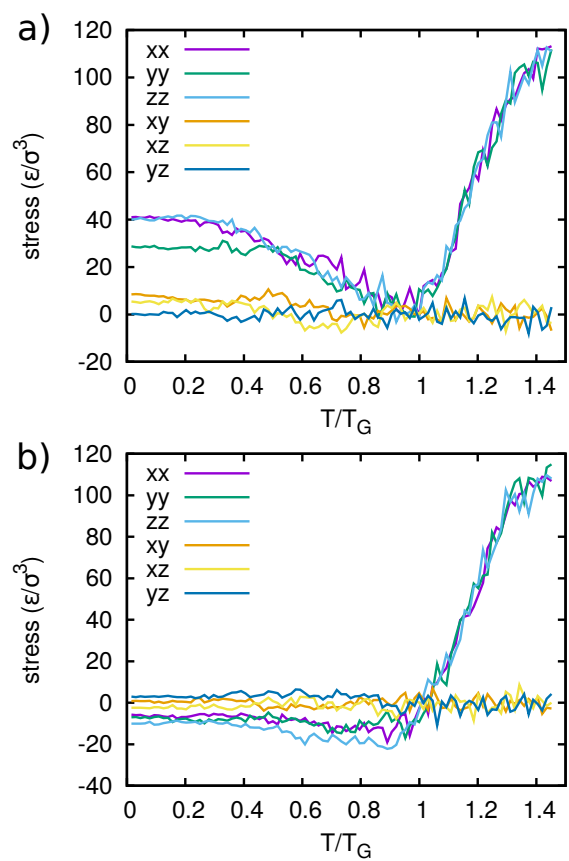

Figure 5: Global stress components as a function of temperature for the CG-q configurations of the a) quenched and b) quench-relaxed samples at the $Q_{-2}$ ramp rate.

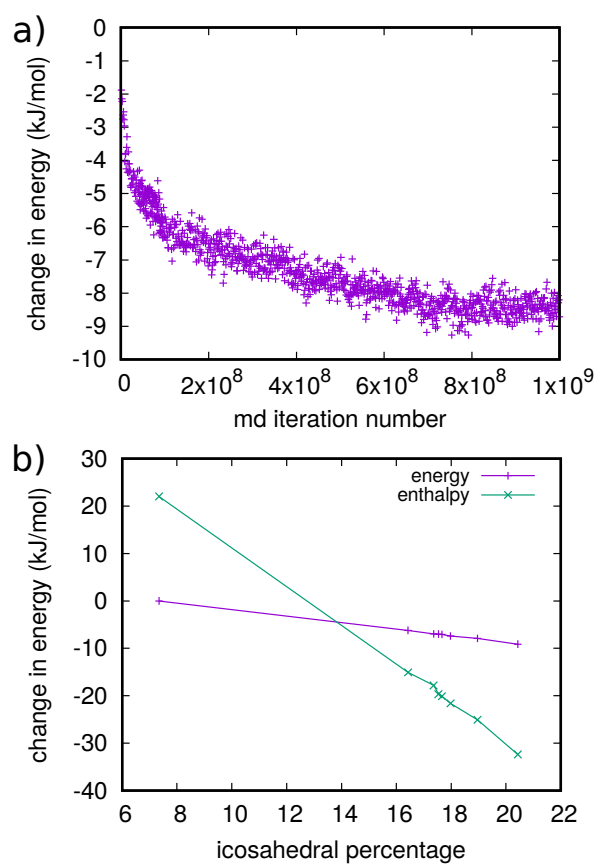

Figure 6: a) Instantaneous potential energy per atom as a function of molecular dynamics iteration number performed at a temperature of $0.95T_G$. b) Zero Kelvin potential energy and enthalpy as a function of icosahedral content. Each data point represents a configuration which has been quenched (at the rate $Q^{LQ}$) from configurations at $T=0.95T_G$, taken at various points along the relaxation curve of a).

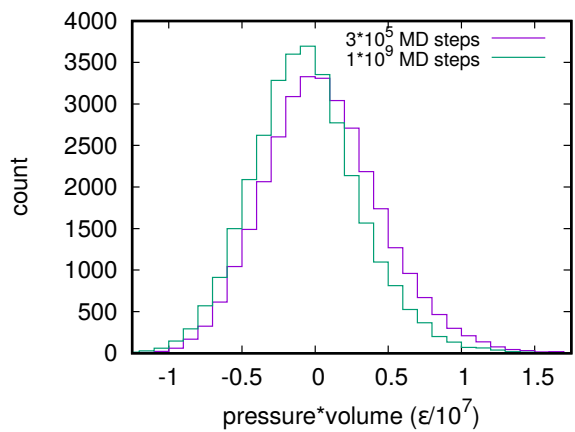

Figure 7: Histograms of local pressure for two quenched samples taken after $10^5$ and $10^9$ molecular dynamics iterations of the relaxation simulation shown in Fig. 6a.